\documentclass[10pt,pra,aps,onecolumn,superscriptaddress]{revtex4-2}
\usepackage{ifthen,amsthm,amsmath,amsxtra,amsfonts,dsfont,graphicx,bm,tikz,scalerel,wasysym,bbm,graphicx,amsthm,braket,physics,empheq}

\usepackage[inline]{enumitem}
\usepackage{CircuitTikz}

\usepackage{hyperref}
\hypersetup{
     colorlinks=true,
     linkcolor=blue,
     filecolor=blue,
     citecolor=blue,      
     urlcolor=blue,
     }
\usepackage{color}

\usepackage{MnSymbol}%
\usepackage{wasysym}%

\newcommand{\fl}{-}

\newcommand{\gateyfs}[1]{
  \begin{tikzpicture}[baseline={([yshift=-0.6ex]current bounding box.center)},scale=0.5]
    \prop{0}{0}{colUinactive}
    \ifthenelse{#1>0}{
      \MEld{0.5}{-#1*0.5}
    }{
      \MErd{0.5}{-#1*0.5}
    }
  \end{tikzpicture}}

\begin{document}

\title{Exact results on the dynamics of the stochastic Floquet-East model}

\author{Cecilia De Fazio}

\affiliation{School of Physics and Astronomy, University of Nottingham, Nottingham, NG7 2RD, UK}
\affiliation{Centre for the Mathematics and Theoretical Physics of Quantum Non-Equilibrium Systems, University of Nottingham, Nottingham, NG7 2RD, UK}

\author{Juan P. Garrahan}

\affiliation{School of Physics and Astronomy, University of Nottingham, Nottingham, NG7 2RD, UK}
\affiliation{Centre for the Mathematics and Theoretical Physics of Quantum Non-Equilibrium Systems, University of Nottingham, Nottingham, NG7 2RD, UK}

\author{Katja Klobas}

\affiliation{School of Physics and Astronomy, University of Nottingham, Nottingham, NG7 2RD, UK}
\affiliation{Centre for the Mathematics and Theoretical Physics of Quantum Non-Equilibrium Systems, University of Nottingham, Nottingham, NG7 2RD, UK}
\affiliation{School of Physics and Astronomy, University of Birmingham, Edgbaston, Birmingham, B15 2TT, UK}

\date{\today}

\begin{abstract}
We introduce a stochastic generalisation of the classical deterministic Floquet-East model, a discrete circuit with the same kinetic constraint as the East model of glasses. We prove exactly that, in the limit of long time and large size, this model has a large deviation phase transition between active and inactive dynamical phases. We also compute the finite time and size scaling of general space-time fluctuations, which for the case of inactive regions gives rise to dynamical hydrophobicity. We also discuss how, through the Trotter limit, these exact results also hold for the continuous-time East model, thus proving long-standing observations in kinetically constrained models. Our results here illustrate the applicability of exact tensor network methods for solving problems in many-body stochastic systems. 
\end{abstract}
\maketitle

\smallskip
\noindent
\emph{This paper is dedicated to the memory of Marko Medenjak (1990--2022).}
\smallskip

\section{Introduction}
In Ref.~\cite{klobas2023exact} we proved via exact methods that the
deterministic Floquet-East (DFE) model exhibits a first-order phase transition in
its dynamical large deviations, and as consequence it displays dynamical
fluctuations corresponding to pre-transition behaviour, a dynamical equivalent
of the hydrophobic effect in
water~\cite{lum1999hydrophobicity,chandler2005interfaces}.  Defined in terms of
a discrete-space/discrete-time circuit with a ``brickwork'' arrangement of
local deterministic (i.e., permutation) gates, the DFE~\cite{klobas2023exact}
(see also~\cite{gopalakrishnan2018facilitated,bernstein2021exotic,bertini2024exact,bertini2024localized})
is a deterministic generalisation (and simplification) of the extensively
studied kinetically constrained East model of
glasses~\cite{jackle1991a-hierarchically,ritort2003glassy}, with the two models
sharing the same kinetic constraint. The standard East model exhibits a phase
transition at the trajectory level where the order parameter is the dynamical
activity, an observable counting the number of configuration
changes~\cite{garrahan2007dynamical,garrahan2009first-order,banuls2019using}.
This is a first-order phase transition between two dynamical phases, an ergodic
equilibrium and dynamically active phase, and a non-ergodic inactive phase.
While the dynamics is at coexistence in the limit of infinite time, for any
finite observation time the large majority of initial conditions favour the
active phase, so that without biasing the typical dynamics of the East model is
ergodic, and the corresponding stationary state is a featureless product state.
Despite that, the existence of the inactive phase has profound consequences:
even in ergodic trajectories mesoscopic fluctuations of inactivity (or
``space-time bubbles'') are predominant, giving rise to dynamic heterogeneity
and glassy relaxation~\cite{garrahan2002geometrical,chandler2010dynamics,garrahan2018aspects}.
These fluctuations are a dynamical analogue of those that give rise to
hydrophobic physics in liquid water~\cite{katira2018solvation}, since the
physical origin is the proximity to a first-order transition (i.e., a dynamical
instance of the {\em orderphobic effect}~\cite{katira2016pre-transition}).

The results of Ref.~\cite{klobas2023exact} for the DFE were obtained via exact
tensor network methods similar to those being currently used to solve certain
cellular automata, see e.g.\
Refs.~\cite{prosen2016integrability,prosen2017exact,gopalakrishnan2018operator,buca2019exact,klobas2019time-dependent,klobas2020space-like,alba2019operator,klobas2020matrix,wilkinson2020exact,iadecola2020nonergodic,gopalakrishnan2018facilitated,gombor2021integrable,wilkinson2022exact},
or  quantum circuits, see e.g.\
Refs.~\cite{bertini2019exact,kos2021correlations,kos2023circuits,bertini2018exact,bertini2019entanglement,bertini2020operator1,bertini2020operator2,piroli2020exact,claeys2020maximum,claeys2021ergodic,bertini2021random,jonay2021triunitary,kasim2022dual,suzuki2022computational,foligno2023growth,foligno2023temporal,rampp2023from,bertini2024localized,bertini2024exact,liu2023solvable}.
The dynamics of the DFE model displays an analogous active-inactive large deviation
transition and corresponding dynamical ``hydrophobic'' fluctuations as the standard
East model, which suggests that the origin of this kind of physics is the local
kinetic constraint that both models share. However, one might wonder if the deterministic
nature of the DFE means that there is a fundamental difference between the two models,
and that the exact results obtained in the DFE are not informative of what
happens when stochasticity is present. In this paper we resolve this question. 

\begin{figure}
\includegraphics[width=18cm]{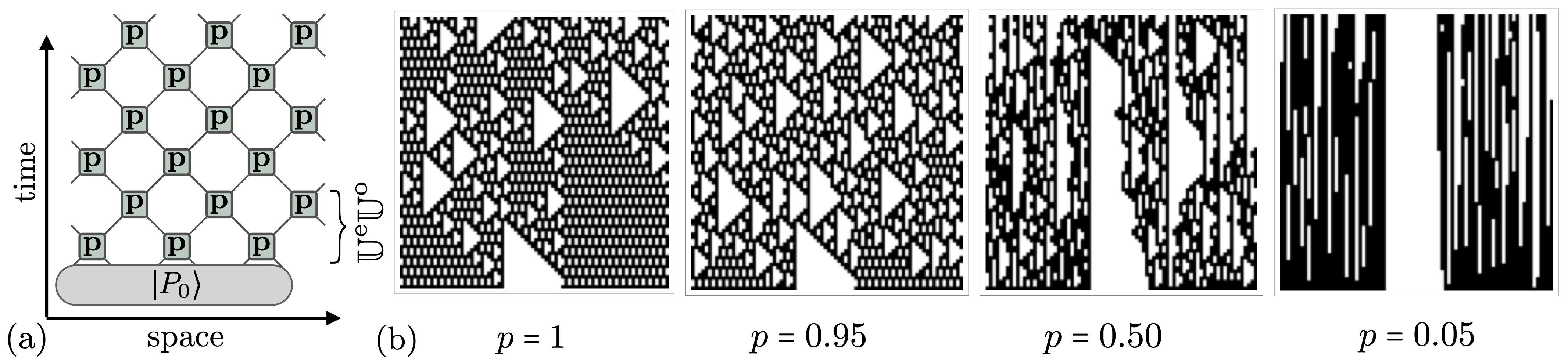}
  \caption{\label{fig:Fig1} Stochastic Floquet-East model.  (a) Diagrammatic
  representation of time-evolution. (b) Examples of trajectories initialised in
  the same domain-wall state
  $\ket{P_0} = \ket{1}^{\otimes 2L/5} \otimes \ket{0}^{\otimes L/5} \otimes \ket{1}^{\otimes 2L/5}$
  for $L=100$ and several values of $p$. States zero/one are represented by white/black dots.}
\end{figure}

We consider a stochastic generalisation of the DFE, or {\em stochastic
Floquet-East} model. Like its deterministic counterpart, the stochastic
Floquet-East is defined as a discrete brickwork circuit (see Fig.~\ref{fig:Fig1}(a) for a diagrammatic representation) with the same East
model kinetic constraint, that is, a site can only flip if its nearest
neighbour ``to the East'' is in the excited state. However, the main difference in the stochastic model is that flips only occur with a
certain probability. Using exact tensor network methods we: (i) prove that in
the long time and large size limit the stochastic Floquet-East also has a large
deviation phase transition between active and inactive dynamical phases; (ii)
compute the finite time and size scaling of space-time fluctuations, showing
that inactive space-time regions gives rise to dynamical hydrophobicity; (iii)
show that the results of the DFE bound those of the
stochastic Floquet-East; and (iv) demonstrate that our results for the
stochastic Floquet-East also hold for the standard continuous-time East model
in an appropriate ``Trotter'' limit, thus proving long-standing observations in
kinetically constrained models. \\

The rest of the paper is organised as follows. In Sec.~\ref{sec:model} we
introduce the model. In Sec.~\ref{sec:LDs}, by studying the large deviations of the activity, we prove the existence of a transition between active and inactive phases, and show that these results also extend to an appropriate limit to the standard continuous-time East model.
In Sec.~\ref{sec:zeros}
we consider the statistics of finite inactive space-time regions, and show
the emergence of dynamical pre-transition (or ``hydrophobic'') effects. For
comparison, in Sec.~\ref{sec:traj} we show that the statistics of arbitrary, non-inactive,
space-time fluctuations. We give our conclusions and outlook in Sec.~\ref{sec:conc}.

\section{Stochastic Floquet-East model}\label{sec:model}
The model is defined on a lattice of $2L$ sites of binary variables
$n_i\in\{0,1\}$ with periodic boundaries, where lattice sites are labelled by half-integer values,
$i\in\{\frac{1}{2},1,\ldots,L-\frac{1}{2},L\}$.  The statistical states $\ket{P}$ are
probability distributions over the finite set of all configurations
$\mathbf{n}=(n_{\frac{1}{2}},n_1,\ldots,n_{L-\frac{1}{2}},n_{L})$,
\begin{equation}
  \ket{P}=\sum_{\mathbf{n}}P(\mathbf{n})\ket{n},\qquad P(\mathbf{n})\ge 0,\qquad
  \sum_{\mathbf{n}}P(\mathbf{n})=\braket{\fl}{P}=1,
\end{equation}
where $ |n\rangle=|n_\frac{1}{2}\rangle\otimes \dots \otimes |n_L\rangle$. The state $\bra{\fl}=\sum_{\mathbf{n}}\bra{\mathbf{n}}$, also called \emph{flat state}, describes the un-normalized uniform distribution. The time-descrete dynamics, shown in Fig.\ref{fig:Fig1}(a), consists of two distinct steps,
\begin{equation}
  \ket*{P_{t+1}}=\mathbb{U}^{\mathrm{e}}\ket*{P_{t+\frac{1}{2}}}=
  \mathbb{U}^{\mathrm{e}} \mathbb{U}^{\mathrm{o}}\ket*{P_{t}},\qquad t\in\mathbb{Z}.
\end{equation}
Note that time is also labelled by half-integer numbers. The time-evolution operators $\mathbb{U}^{\mathrm{o/e}}$ consist of mutually
commuting nearest-neighbour updates,
\begin{equation}
  \mathbb{U}^{\mathrm{e}}=U_p^{\otimes L},\qquad
  \mathbb{U}^{\mathrm{o}}=\Pi_L U_p^{\otimes L} \Pi_L^{\dagger},
\end{equation}
where $\Pi_L$ is a one-site shift operator, and $U_p$ the $4\times 4$ matrix
implementing the two-site update
\begin{equation}
  \mel{m^{\prime} n^{\prime}}{U_p}{m n}=\delta_{n^{\prime},0}\delta_{n,0} \delta_{m^{\prime},m}
  +\delta_{n^{\prime},1}\delta_{n,1}
  \left[\delta_{m^{\prime},1-m} p +\delta_{m^{\prime},m}(1-p)\right],
\end{equation}
with $|m n\rangle=| m \rangle \otimes | n \rangle$.
The above matrix encodes the kinetic constraint: if the right site is in state one, the spin on the left site changes with fixed probability $0\le p\le 1$, and stays the same otherwise.  \\ 

Throughout the paper we will make use of the diagrammatic tensor network
notation~\cite{tensor}, where tensors are represented by nodes (or vertices)
with lines emanating from them. The number of lines corresponds to the rank of
the tensor, and lines connecting two nodes indicating a contraction along that
dimension. For example, in this representation local vectors corresponding to
one physical site are
\begin{equation} 
  \begin{tikzpicture}[baseline={([yshift=-0.6ex]current bounding box.center)},scale=0.5]
    \nctgridLine{0}{0}{0}{0.75}
    \draw[thick,colLines,fill=white] (0,0) circle (0.25);
    \node at (0,0) {\scalebox{0.8}{$v$}};
  \end{tikzpicture} =
  \ket{v}=v_0\ket{0}+v_1\ket{1},\qquad
  \begin{tikzpicture}[baseline={([yshift=-0.6ex]current bounding box.center)},scale=0.5]
    \nctgridLine{0}{0}{0}{0.75}
    \node[draw=white,fill=white,circle,inner sep=0] at (0,-0.25) {\scalebox{0.8}{$0$}};
  \end{tikzpicture} =\ket{0},\qquad
  \begin{tikzpicture}[baseline={([yshift=-0.6ex]current bounding box.center)},scale=0.5]
    \nctgridLine{0}{0}{0}{0.75}
    \node[draw=white,fill=white,circle,inner sep=0] at (0,-0.25) {\scalebox{0.8}{$1$}};
  \end{tikzpicture} =\ket{1},\qquad
  \begin{tikzpicture}[baseline={([yshift=-0.6ex]current bounding box.center)},scale=0.5]
    \nctgridLine{0}{0}{0}{0.75}
    \MEh{0}{0}
  \end{tikzpicture} =
  \begin{tikzpicture}[baseline={([yshift=-0.6ex]current bounding box.center)},scale=0.5]
    \nctgridLine{0}{0}{0}{0.75}
    \node[draw=white,fill=white,circle,inner sep=0] at (0,-0.25) {\scalebox{0.8}{$0$}};
  \end{tikzpicture}+
  \begin{tikzpicture}[baseline={([yshift=-0.6ex]current bounding box.center)},scale=0.5]
    \nctgridLine{0}{0}{0}{0.75}
    \node[draw=white,fill=white,circle,inner sep=0] at (0,-0.25) {\scalebox{0.8}{$1$}};
  \end{tikzpicture} = \ket{\fl},
\end{equation}
where we introduced a shorthand notation for computational-basis states, and
the one-site flat state $\ket{\fl}$. In this language a free line represents a
one-site identity matrix, and the inner product is given by contracting two
vectors,
\begin{equation}
  \braket{w}{v}=
  \begin{tikzpicture}[baseline={([yshift=-0.6ex]current bounding box.center)},scale=0.5]
    \nctgridLine{0}{0}{0}{1}
    \draw[thick,colLines,fill=white] (0,0) circle (0.25);
    \draw[thick,colLines,fill=white] (0,1) circle (0.25);
    \node at (0,0) {\scalebox{0.8}{$v$}};
    \node at (0,1) {\scalebox{0.8}{$w$}};
  \end{tikzpicture},\qquad
  \begin{tikzpicture}[baseline={([yshift=-0.6ex]current bounding box.center)},scale=0.5]
    \nctgridLine{0}{0}{0}{1}
    \end{tikzpicture}=\begin{bmatrix} 1&0\\0&1\end{bmatrix}.
\end{equation}
We introduce a green box with $4$ legs to represent the tensor which, when
interpreted as acting upwards, gives $U_p$,
\begin{equation} 
  \begin{tikzpicture}[baseline={([yshift=-0.85ex]current bounding box.center)},scale=0.5]
    \propS{0}{0}{colU}{black}
    \node at (-0.875,-0.875) {$m$};
    \node at (0.875,-0.875) {$n$};
    \node at (-0.875,0.875) {$m^{\prime}$};
    \node at (0.875,0.875) {$n^{\prime}$};
  \end{tikzpicture}\mkern-8mu =
  \mel{m^{\prime}n^{\prime}}{U_p}{mn}.
\end{equation}
Using these conventions, the full time-evolution can be represented as the tensor network shown
in Fig.~\ref{fig:Fig1}(a). In the limiting cases of $p=1$ or $p=0$ the dynamics becomes deterministic,
corresponding to the DFE 
model~\cite{klobas2023exact,bertini2024exact} and trivial (identity) dynamics, 
respectively,
\begin{equation}
  \begin{tikzpicture}[baseline={([yshift=-0.85ex]current bounding box.center)},scale=0.5]
    \propSOne{0}{0}{colU}{black}
  \end{tikzpicture}=:
  \begin{tikzpicture}[baseline={([yshift=-0.85ex]current bounding box.center)},scale=0.5]
    \prop{0}{0}{colU}
  \end{tikzpicture},\qquad
  \begin{tikzpicture}[baseline={([yshift=-0.85ex]current bounding box.center)},scale=0.5]
    \propSZero{0}{0}{colU}{black}
  \end{tikzpicture}=
  \begin{tikzpicture}[baseline={([yshift=-0.6ex]current bounding box.center)},scale=0.5]
    \bendRud{0}{-0.5}{0.5}
    \bendLud{0.75}{-0.5}{0.5}
  \end{tikzpicture}.
\end{equation}
Fig.\ref{fig:Fig1}(b) shows some typical trajectories for systems prepared in the same domain-wall initial state and various probability $p$. For $p=1$, trajectories are characterised by the presence of stationary space-time regions of triangular shape. These empty regions can be still observed for $0<p<1$, however, they are not necessarily triangular, and may be much more extended in time. \\

  For later convenience, we define the \emph{tilted} gate $U_{p,s}$, where each time
  the spin-flip occurs, we introduce a factor of $\mathrm{e}^{-s}$,
  which we represent by a dark-shaded box,
  \begin{equation}\label{eq:defTilted}
  U_{p,s}=\begin{bmatrix}
    1& & & \\
    &1-p&&\mathrm{e}^{-s}p\\
    &&1&\\
    &\mathrm{e}^{-s}p& &1-p
  \end{bmatrix},\qquad
  \begin{tikzpicture}[baseline={([yshift=-0.85ex]current bounding box.center)},scale=0.5]
    \propS{0}{0}{colUtilted}{white}
    \node at (-0.875,-0.875) {$m$};
    \node at (0.875,-0.875) {$n$};
    \node at (-0.875,0.875) {$m^{\prime}$};
    \node at (0.875,0.875) {$n^{\prime}$};
  \end{tikzpicture}\mkern-8mu =
  \mel{m^{\prime}n^{\prime}}{U_{p,s}}{mn},\qquad
  \begin{tikzpicture}[baseline={([yshift=-0.6ex]current bounding box.center)},scale=0.5]
    \propS{0}{0}{colUtilted}{white}
  \end{tikzpicture}=
  \begin{tikzpicture}[baseline={([yshift=-0.6ex]current bounding box.center)},scale=0.5]
    \propS{0}{0}{colUinactive}{black}
  \end{tikzpicture}
  +
  e^{-s}
  \begin{tikzpicture}[baseline={([yshift=-0.6ex]current bounding box.center)},scale=0.5]
    \propS{0}{0}{colUactive}{black}
  \end{tikzpicture}.
\end{equation}
Here, the blue and red squares are the \emph{inactive} and \emph{active} parts of the gate (i.e., they include the matrix elements with no-flips and flips respectively). By inspecting their matrix elements, we immediately note that the inactive gate factorizes into an identity on the left, and a diagonal transformation on the right site,
\begin{equation}\label{eq:defInactiveP}
  \begin{tikzpicture}[baseline={([yshift=-0.6ex]current bounding box.center)},scale=0.5]
    \propS{0}{0}{colUinactive}{black}
  \end{tikzpicture}=
  \begin{tikzpicture}[baseline={([yshift=-0.6ex]current bounding box.center)},scale=0.5]
    \bendRud{0}{-0.5}{0.5}
    \bendLud{0.75}{-0.5}{0.5}
    \obsWiggly{0.55}{0}
  \end{tikzpicture},\qquad
  \begin{tikzpicture}[baseline={([yshift=-0.6ex]current bounding box.center)},scale=0.5]
    \tgridLine{0}{-0.5}{0}{0.5}
    \obsWiggly{0}{0}
  \end{tikzpicture}=
  \begin{bmatrix}
    1 & 0 \\ 
    0 & 1-p
  \end{bmatrix},
\end{equation}
while the active gate is the same as the active gate in the $p=1$ case
multiplied by a factor of $p$,
\begin{equation}\label{eq:defActiveP}
  \begin{tikzpicture}[baseline={([yshift=-0.6ex]current bounding box.center)},scale=0.5]
    \propS{0}{0}{colUactive}{black}
  \end{tikzpicture}=
  p\ 
  \begin{tikzpicture}[baseline={([yshift=-0.6ex]current bounding box.center)},scale=0.5]
    \prop{0}{0}{colUactive}
  \end{tikzpicture},\qquad
  \begin{tikzpicture}[baseline={([yshift=-0.85ex]current bounding box.center)},scale=0.5]
    \prop{0}{0}{colUactive}
    \node at (-0.875,-0.875) {$m$};
    \node at (0.875,-0.875) {$n$};
    \node at (-0.875,0.875) {$m^{\prime}$};
    \node at (0.875,0.875) {$n^{\prime}$};
  \end{tikzpicture}=\delta_{m,1-m^{\prime}}\delta_{n,1}\delta_{n^{\prime},1}.
\end{equation}
Apart from the diagonal transformation introduced in~\eqref{eq:defInactiveP}, we will make use two other one-site observables: a projector to the zero and one state, respectively represented by a white and black circle,
\begin{equation}
  \begin{tikzpicture}[baseline={([yshift=-0.6ex]current bounding box.center)},scale=0.5]
    \tgridLine{0}{-0.5}{0}{0.5}
    \obsZero{0}{0}
  \end{tikzpicture}=
  \begin{bmatrix}
    1 & 0 \\ 
    0 & 0
  \end{bmatrix},\qquad
  \begin{tikzpicture}[baseline={([yshift=-0.6ex]current bounding box.center)},scale=0.5]
    \tgridLine{0}{-0.5}{0}{0.5}
    \obsOne{0}{0}
  \end{tikzpicture}=
  \begin{bmatrix}
    0 & 0 \\ 
    0 & 1
  \end{bmatrix}.
\end{equation}

\begin{figure}
\includegraphics[width=0.75\textwidth]{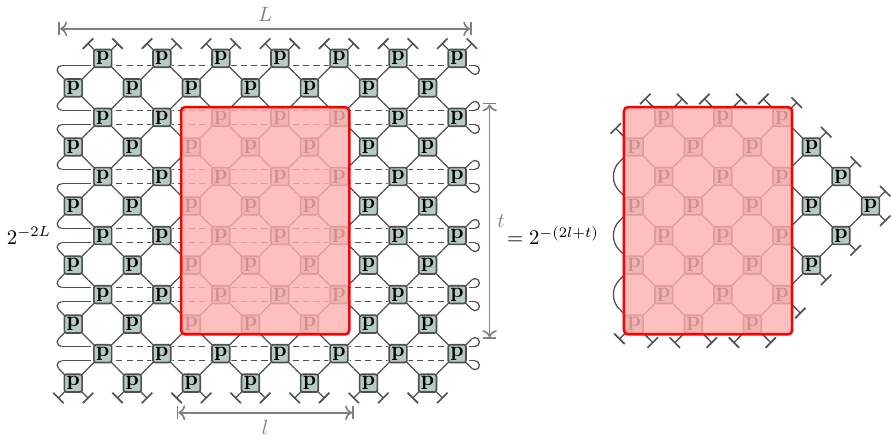}
  \caption{\label{fig:Fig2} Local correlations and expectation values in a space-time box embedded in a large system. The l.h.s.\ shows a schematic representation of an infinite-temperature correlation function between observables that are confined to a space-time box of sizes $l\times t$ (represented by the red rectangle), embedded in a larger system of size $L$ with periodic boundary conditions.  As long as $l+t<L$, we can repeatedly apply local relations in Eqs.~\eqref{eq:bistoch},~\eqref{eq:spaceEvol} to recast the object as the same correlation function (represented by the red box), but now evaluated in a finite system with special boundary conditions (r.h.s.). Note that the size of the green rectangle on the r.h.s.\ scales with time $t$.}
\end{figure}

The model is bi-stochastic, that is, both the dynamics and its inverse conserve
probability, which is a consequence of the flat state being both a left and
right eigenvector of the local time-evolution operator
\begin{equation}
  U_{p}\ket{\fl}\otimes\ket{\fl}=\ket{\fl}\otimes\ket{\fl},\qquad
  \bra{\fl}\otimes\bra{\fl}U_p=\bra{\fl}\otimes\bra{\fl}.
\end{equation}
These two local relations can be equivalently represented using the above graphical language as
\begin{equation}\label{eq:bistoch}
  \begin{tikzpicture}[baseline={([yshift=-0.85ex]current bounding box.center)},scale=0.5]
    \propS{0}{0}{colU}{black}
    \MErd{0.5}{0.5}
    \MEld{-0.5}{0.5}
  \end{tikzpicture}=
  \begin{tikzpicture}[baseline={([yshift=-0.85ex]current bounding box.center)},scale=0.5]
    \bendRud{-0.25}{-0.5}{0.5}
    \bendLud{0.5}{-0.5}{0.5}
    \MEld{-0.25}{0.5}
    \MErd{0.5}{0.5}
  \end{tikzpicture},\qquad
  \begin{tikzpicture}[baseline={([yshift=-0.45ex]current bounding box.center)},scale=0.5]
    \propS{0}{0}{colU}{black}
    \MErd{-0.5}{-0.5}
    \MEld{0.5}{-0.5}
  \end{tikzpicture}=
  \begin{tikzpicture}[baseline={([yshift=-0.45ex]current bounding box.center)},scale=0.5]
    \bendRud{-0.25}{-0.5}{0.5}
    \bendLud{0.5}{-0.5}{0.5}
    \MEld{0.5}{-0.5}
    \MErd{-0.25}{-0.5}
  \end{tikzpicture}.
\end{equation}
Moreover, the gates exhibit a stronger property: whenever one acts with a flat state from either top or bottom on the \emph{left} site, it factorizes into a projector to the flat state and an identity operator,
\begin{equation}\label{eq:spaceEvol}
  \begin{tikzpicture}[baseline={([yshift=-0.85ex]current bounding box.center)},scale=0.5]
    \propS{0}{0}{colU}{black}
    \MEld{-0.5}{0.5}
  \end{tikzpicture}=
  \begin{tikzpicture}[baseline={([yshift=-0.85ex]current bounding box.center)},scale=0.5]
    \bendRud{-0.25}{-0.5}{0.5}
    \bendLud{0.5}{-0.5}{0.5}
    \MEld{-0.25}{0.5}
  \end{tikzpicture},\qquad
  \begin{tikzpicture}[baseline={([yshift=-0.45ex]current bounding box.center)},scale=0.5]
    \propS{0}{0}{colU}{black}
    \MErd{-0.5}{-0.5}
  \end{tikzpicture}=
  \begin{tikzpicture}[baseline={([yshift=-0.45ex]current bounding box.center)},scale=0.5]
    \bendRud{-0.25}{-0.5}{0.5}
    \bendLud{0.5}{-0.5}{0.5}
    \MErd{-0.25}{-0.5}
  \end{tikzpicture},
\end{equation}
which can be understood as a stochastic analogue of the generalization of dual
unitarity introduced in Ref.~\cite{yu2024hierarchical}, but here it only
applies to one side. This property implies that an invariant state of the
space-evolution coming from the left consists of dimer product states, as shown
in Fig.~\ref{fig:Fig2}. We note that for $p=1$ the gates preserve
flat states coming from the right, which implies a simple form also for the
invariant states under space evolution from right to
left~\cite{klobas2023exact}, but this ceases to hold for generic $p$.

\section{Dynamical large deviations}\label{sec:LDs}

As is proven in Ref.~\cite{klobas2023exact}, the DFE
model exhibits a first-order phase transition between and ``active'' and
an ``inactive'' dynamical phase -- phases with a large and small number of
spin flips, respectively. Below, we show that the same is true for the stochastic
generalisation. \\

We consider a space-time region of size $l\times t$ in a much larger system of
size $L\times T$, and we take the dynamical order parameter to be the dynamical
activity -- total number of spin flips in the region $l\times t$. Let
$\{\omega\}$ be the set of all the possible trajectories in the full system,
$\pi_p(\omega)$ probability of the trajectory $\omega$, and $K_{l,t}(\omega)$ its
activity. Then the moment generating function of the activity is given by 
\cite{lecomte2007thermodynamic,touchette2009the-large,jack2020ergodicity}
\begin{equation}
  Z_{l,t}(p,s)=\sum_{\omega} \pi_p(\omega) \mathrm{e}^{-s K_{l,t}(\omega)},
\end{equation}
where we will assume that the initial configuration is uniformly distributed.
Equivalently, $Z_{l,t}(p,s)$ can be interpreted as a partition function for a
\emph{tilted region} [i.e., a region where the stochastic gates $U_p$ are
replaced by tilted gates~\eqref{eq:defTilted}] of size $l\times t$ embedded in
a much larger system of size $L\times T$, which can be diagrammatically
represented as 
\begin{equation} \label{eq:defPartitionFunction}
  Z_{l,t}(s)= 2^{-(2l+t)}
  \begin{tikzpicture}[baseline={([yshift=-0.6ex]current bounding box.center)},scale=0.5]
    \draw[thick,|<->|,gray] (2,1) -- (8,1) node[midway,below] {$l$};
    \draw[thick,|<->|,gray] (1.5,1.5) -- (1.5,9.5) node[midway,left] {$t$};
    \foreach \x in {2,4,6}{
      \foreach \y in {2,4,6,8}{
        \propS{\x+0.5}{\y}{colUtilted}{white}
        \propS{\x+1.5}{\y+1}{colUtilted}{white}
      }
    }
    \foreach \x in {2,4,6}{
      \MErd{\x}{1.5}
      \MEld{\x+1}{1.5}
      \MEld{\x+1}{9.5}
      \MErd{\x+2}{9.5}
    }
    \foreach \t in {2,4,6}{\bendLud{2}{\t+0.5}{\t+1.5}}
    \MEld{2}{8.5}
    \foreach \t in {4,6,8}{\propS{8.5}{\t}{colU}{black}}
    \foreach \t in {5,7}{\propS{9.5}{\t}{colU}{black}}
    \foreach \t in {6}{\propS{10.5}{\t}{colU}{black}}
    \foreach \t in {3,...,6}{\MEld{5+\t}{\t-0.5}}
    \foreach \t in {7,...,9}{\MErd{18-\t}{\t-0.5}}
  \end{tikzpicture}.
\end{equation}
The contraction that leads from the large space-time region $L \times T$ to the above diagram has been obtained by using relations \eqref{eq:bistoch} and \eqref{eq:spaceEvol} and by following the procedure described in Fig.~\ref{fig:Fig2}. Though conceptually simple, the diagram in Eq.~\eqref{eq:defPartitionFunction} cannot be evaluated exactly for all $s$. However, we are able to bound the behaviour effectively in two limits,
$s\approx 0$, and $s\to\infty$, which -- upon evoking general properties of
the moment generating function $F_{l,t}(p,s)=\log Z_{l,t}(p,s)$ -- suffices to
demonstrate the existence of a first-order phase transition.

\subsection{Expansion around $s\approx 0$}\label{sec:Zat0}
The first regime of interest is $s\approx 0$, where $Z_{l,t}(p,s)$ is well approximated by the expansion up to $s^2$,
\begin{equation}
  Z_{l,t}(p,s)\approx 1 - s Z_{l,t}^{(1)}(p)+ s^2 Z_{l,t}^{(2)}(p)+\mathcal{O}(s^3),
\end{equation}
where we took into account that the bi-stochasticity~\eqref{eq:bistoch} implies
$Z_{l,t}(p,0)=1$. By expanding each of the tilted gates as
\begin{equation}
  \begin{tikzpicture}[baseline={([yshift=-0.6ex]current bounding box.center)},scale=0.5]
    \propS{5.5}{3}{colUtilted}{white}
  \end{tikzpicture}=
  \begin{tikzpicture}[baseline={([yshift=-0.6ex]current bounding box.center)},scale=0.5]
    \propS{5.5}{3}{colU}{black}
  \end{tikzpicture}+
  \left(\mathrm{e}^{-s}-1\right)
  \begin{tikzpicture}[baseline={([yshift=-0.6ex]current bounding box.center)},scale=0.5]
    \propS{5.5}{3}{colUactive}{black}
  \end{tikzpicture},
\end{equation}
one sees that in the first-order the contribution is given in terms of one-point functions
in the stationary state, which can be easily evaluated as
\begin{equation}
  Z_{l,t}^{(1)}(p)=
  \sum_{\{x,y\}}
    2^{-(2l+t)}
    \begin{tikzpicture}[baseline={([yshift=-0.6ex]current bounding box.center)},scale=0.5]
      \foreach \x in {2,4,6}{
        \foreach \y in {2,4,6,8}{
          \propS{\x+0.5}{\y}{colU}{black}
          \propS{\x+1.5}{\y+1}{colU}{black}
        }
      }
      \propS{5.5}{3}{colUactive}{black}
      \foreach \x in {2,4,6}{
        \MErd{\x}{1.5}
        \MEld{\x+1}{1.5}
        \MEld{\x+1}{9.5}
        \MErd{\x+2}{9.5}
      }
      \foreach \t in {2,4,6}{\bendLud{2}{\t+0.5}{\t+1.5}}
      \MEld{2}{8.5}
      \foreach \t in {2,3,4,5}{\MEld{6+\t}{0.5+\t}}
      \foreach \t in {6,7,8}{\MErd{17-\t}{0.5+\t}}
      \foreach \t in {4,6,8}{\propS{8.5}{\t}{colU}{black}}
      \foreach \t in {5,7}{\propS{9.5}{\t}{colU}{black}}
      \foreach \t in {6}{\propS{10.5}{\t}{colU}{black}}
      \draw [thick,gray,|<->|] (2.5,1) -- (5.5,1) node[midway,below]{$x$};
      \draw [thick,gray,|<->|] (1.5,2) -- (1.5,3) node[midway,left]{$y$};
    \end{tikzpicture}=
    \sum_{\{x,y\}}
    \frac{1}{4}
  \begin{tikzpicture}[baseline={([yshift=-0.6ex]current bounding box.center)},scale=0.5]
    \propS{5.5}{3}{colUactive}{black}
    \MErd{5}{2.5}
    \MEld{5}{3.5}
    \MEld{6}{2.5}
    \MErd{6}{3.5}
  \end{tikzpicture}= p l t,
\end{equation}
where the sum goes over all positions $\{x,y\}$ inside the rectangular part [i.e., the shaded part in~\eqref{eq:defPartitionFunction}]. The first equality above, again, follows directly from the bi-stochasticity, and the final result from the definition of the gate. The second-order correction requires some more work, as now we get
contributions both from the one and two-point functions,
\begin{equation}
  Z_{l,t}^{(2)}(p)=\frac{1}{2} Z_{l,t}^{(1)}(p)
  +\frac{1}{2}\smashoperator[r]{\sum_{\{x_1,y_1\}\neq\{x_2,y_2\}}}
  \underbrace{
    p^2
    2^{-(2l+t)}
    \begin{tikzpicture}[baseline={([yshift=-0.6ex]current bounding box.center)},scale=0.5]
      \foreach \x in {2,4,6}{
        \foreach \y in {2,4,6,8}{
          \propS{\x+0.5}{\y}{colU}{black}
          \propS{\x+1.5}{\y+1}{colU}{black}
        }
      }
      \prop{5.5}{3}{colUactive}
      \prop{3.5}{7}{colUactive}
      \foreach \x in {2,4,6}{
        \MErd{\x}{1.5}
        \MEld{\x+1}{1.5}
        \MEld{\x+1}{9.5}
        \MErd{\x+2}{9.5}
      }
      \foreach \t in {2,4,6}{\bendLud{2}{\t+0.5}{\t+1.5}}
      \MEld{2}{8.5}
      \foreach \t in {2,3,4,5}{\MEld{6+\t}{0.5+\t}}
      \foreach \t in {6,7,8}{\MErd{17-\t}{0.5+\t}}
      \foreach \t in {4,6,8}{\propS{8.5}{\t}{colU}{black}}
      \foreach \t in {5,7}{\propS{9.5}{\t}{colU}{black}}
      \foreach \t in {6}{\propS{10.5}{\t}{colU}{black}}
      \draw [thick,gray,|<->|] (2.5,1) -- (5.5,1) node[midway,below]{$x_1$};
      \draw [thick,gray,|<->|] (8.5,2) -- (8.5,3) node[midway,right]{$y_1$};
      \draw [thick,gray,|<->|] (2.5,10) -- (3.5,10) node[midway,above]{$x_2$};
      \draw [thick,gray,|<->|] (1.5,2) -- (1.5,7) node[midway,left]{$y_2$};
    \end{tikzpicture}}_{C_{2}(\{x_1,y_1\},\{x_2,y_2\};p)}.
\end{equation}
Using the observation
\begin{equation}
  \begin{tikzpicture}[baseline={([yshift=-0.6ex]current bounding box.center)},scale=0.5]
    \propS{0}{0}{colU}{black}
    \MEld{0.5}{-0.5}
    \MEld{-0.5}{0.5}
  \end{tikzpicture}=
  \begin{tikzpicture}[baseline={([yshift=-0.6ex]current bounding box.center)},scale=0.5]
    \tgridLine{-0.125}{-0.125}{-0.5}{-0.5}
    \tgridLine{0.125}{0.125}{0.5}{0.5}
    \MErd{-0.125}{-0.125}
    \MErd{0.125}{0.125}
  \end{tikzpicture},\qquad
  \begin{tikzpicture}[baseline={([yshift=-0.6ex]current bounding box.center)},scale=0.5]
    \propS{0}{0}{colU}{black}
    \MErd{-0.5}{-0.5}
    \MErd{0.5}{0.5}
  \end{tikzpicture}=
  \begin{tikzpicture}[baseline={([yshift=-0.6ex]current bounding box.center)},scale=0.5]
    \tgridLine{0.125}{-0.125}{0.5}{-0.5}
    \tgridLine{-0.125}{0.125}{-0.5}{0.5}
    \MEld{0.125}{-0.125}
    \MEld{-0.125}{0.125}
  \end{tikzpicture},
\end{equation}
and explicitly treating the case $\{x_2,y_2\}=\{x_1,y_1\}+\{\pm 1,\pm 1\}$, one can show 
that whenever $x_1\neq x_2$ the correlation function factorizes,
\begin{equation}
  \left.C_2(\{x_1,y_1\},\{x_2,y_2\};p)\right|_{x_2\neq x_1}=
  \left(\frac{1}{4}
  \begin{tikzpicture}[baseline={([yshift=-0.6ex]current bounding box.center)},scale=0.5]
    \propS{0}{0}{colUactive}{black}
    \MErd{0.5}{0.5}
    \MErd{-0.5}{-0.5}
    \MEld{-0.5}{0.5}
    \MEld{0.5}{-0.5}
  \end{tikzpicture}
  \right)^2=\frac{p^2}{4}.
\end{equation}
When $x_1=x_2$, however, the two-point correlation function no longer
factorizes, but instead depends on the distance between the two points,
$|y_2-y_1|$, which can take only positive integer values between $1$ and $t-1$,
\begin{equation}
  C_2(\{x,y\},\{x,y\pm m\};p)=
  2^{-(m+2)}
  \begin{tikzpicture}[baseline={([yshift=-0.6ex]current bounding box.center)},scale=0.5]
    \foreach \t in {3,5,7}{\bendLud{7}{\t+0.5}{\t+1.5}}
    \foreach \t in {3,9}{\propS{7.5}{\t}{colUactive}{black}}
    \foreach \t in {5,7}{\propS{7.5}{\t}{colU}{black}}
    \foreach \t in {4,6,8}{\propS{8.5}{\t}{colU}{black}}
    \foreach \t in {5,7}{\propS{9.5}{\t}{colU}{black}}
    \foreach \t in {6}{\propS{10.5}{\t}{colU}{black}}
    \foreach \t in {3,...,6}{\MEld{5+\t}{\t-0.5}}
    \foreach \t in {7,...,10}{\MErd{18-\t}{\t-0.5}}
    \MEld{7}{9.5}
    \MErd{7}{2.5}
    \draw [thick,gray,|<->|] (6.5,3) -- (6.5,9) node[midway,left]{$m$};
  \end{tikzpicture}=
  p^2 2^{-(m+1)}
  \begin{tikzpicture}[baseline={([yshift=-0.6ex]current bounding box.center)},scale=0.5]
    \foreach \t in {4,6}{\bendLud{8}{\t+0.5}{\t+1.5}}
    \foreach \t in {4,6,8}{\propS{8.5}{\t}{colU}{black}}
    \foreach \t in {5,7}{\propS{9.5}{\t}{colU}{black}}
    \foreach \t in {6}{\propS{10.5}{\t}{colU}{black}}
    \foreach \t in {4,...,6}{\MEld{5+\t}{\t-0.5}}
    \foreach \t in {7,...,9}{\MErd{18-\t}{\t-0.5}}
    \tgridLine{8}{8.5}{7.75}{8.75}
    \tgridLine{8}{3.5}{7.75}{3.25}
    \obsOne{8}{8.5}
    \obsOne{8}{3.5}
    \MEld{7.75}{8.75}
    \MErd{7.75}{3.25}
    \draw [thick,gray,|<->|] (7.5,4) -- (7.5,8) node[midway,left]{$m-1$};
  \end{tikzpicture}=
  p^2 c_2(p,m-1),
\end{equation}
where we introduced the short-hand notation $c_2(p,m)$ in terms of which the full second-order contribution reads as
\begin{equation}
  Z_{l,t}^{(2)}(p)=\frac{p l t}{2}+\frac{p^2}{4} l t (2lt-1)
  + 2 l p^2 \sum_{m=0}^{t-2} \left(c_2(p,m)-\frac{1}{4}\right).
\end{equation}
We are not able to evaluate $c_2(p,m)$ exactly, however, we are able to provide some bounds on its magnitude. We start by noting that it can be understood as a matrix element
\begin{equation}
  c_2(p,m)=\frac{1}{2}
  \begin{bmatrix} 0 & 1\end{bmatrix} R_m(p) \begin{bmatrix} 0 \\ 1 \end{bmatrix},\qquad
    R_m(p)=
  2^{-(m+1)}
  \begin{tikzpicture}[baseline={([yshift=-0.6ex]current bounding box.center)},scale=0.5]
    \foreach \t in {4,6}{\bendLud{8}{\t+0.5}{\t+1.5}}
    \foreach \t in {4,6,8}{\propS{8.5}{\t}{colU}{black}}
    \foreach \t in {5,7}{\propS{9.5}{\t}{colU}{black}}
    \foreach \t in {6}{\propS{10.5}{\t}{colU}{black}}
    \foreach \t in {4,...,6}{\MEld{5+\t}{\t-0.5}}
    \foreach \t in {7,...,9}{\MErd{18-\t}{\t-0.5}}
    \draw [thick,gray,|<->|] (7.5,4) -- (7.5,8) node[midway,left]{$m$};
  \end{tikzpicture},
\end{equation}
where $R_m(p)$ denotes the matrix between the free legs on bottom and
top. As individual gates comprising $R_m(p)$ are bi-stochastic matrices, also
$R_m(p)$ is bi-stochastic, therefore there exists a $m$ and $p$-dependent value
$r_m(p)$ so that
\begin{equation}
  0\le r_m(p) \le 1,\qquad
  R_m(p)=\begin{bmatrix}
    r_m& 1-r_m\\
    1-r_m & r_m
  \end{bmatrix}.
\end{equation}
Expressing $c_2(p,m)$ in terms of $r_m$ we obtain both a lower and an upper bound,
\begin{equation}
  0\le c_2(p,m)=\frac{1}{2} r_m \le \frac{1}{2},
\end{equation}
with the lower bound $c_2(p,m)\ge 0$ being trivial, as it follows directly from
non-negativity of the matrix elementes. However, the upper bound is tight,
since it is saturated for $p=0$,
\begin{equation}
  c_2(p,m)\le \frac{1}{2}=c_2(0,m).
\end{equation}

Using these, we can bound the second order contribution as
\begin{equation} \label{eq:boundSecondOrderZ}
  \frac{(p l t)^2}{2}+\frac{p l t}{4}\left(2-p(3-\frac{2}{t})\right)
  \le
  Z_{l,t}^{(2)}(p)
  \le 
  \frac{(p l t)^2}{2}+\frac{p l t}{4}\left(2+p(1-\frac{2}{t})\right).
\end{equation}
This is enough to understand the asymptotics of $Z_{l,t}^{(2)}$: in the limit of $l,t\to\infty$ with $l/t$ fixed, the leading order contribution is $(p l t)^2/2$, with the precise form of $c_2(p,m)$ determining the subleading term.

\subsection{$s\to\infty$ limit}
Let us proceed by considering the $s\to\infty$ limit of $Z_{l,t}(p,s)$.  We
start by noticing that the stochastic generalized inactive gate, when seen as
evolving in time, acts on the two sites as a product of an identity and a
diagonal transformation respectively, as shown in Eq.~\eqref{eq:defInactiveP}.
Using this representation of the gate we can express the partition sum as
\begin{equation}
  \begin{aligned}
    \lim_{s\to\infty} Z_{l,t}(p,s)&=
  2^{-(2l+t)}
  \begin{tikzpicture}[baseline={([yshift=-0.6ex]current bounding box.center)},scale=0.5]
    \foreach \x in {2,4,6}{
      \foreach \y in {2,4,6,8}{
        \propS{\x+0.5}{\y}{colUinactive}{black}
        \propS{\x+1.5}{\y+1}{colUinactive}{black}
      }
    }
    \foreach \x in {2,4,6}{
      \MErd{\x}{1.5}
      \MEld{\x+1}{1.5}
      \MEld{\x+1}{9.5}
      \MErd{\x+2}{9.5}
    }
    \foreach \t in {2,4,6}{\bendLud{2}{\t+0.5}{\t+1.5}}
    \MEld{2}{8.5}
    \foreach \t in {4,6,8}{\propS{8.5}{\t}{colU}{black}}
    \foreach \t in {5,7}{\propS{9.5}{\t}{colU}{black}}
    \foreach \t in {6}{\propS{10.5}{\t}{colU}{black}}
    \foreach \t in {3,...,6}{\MEld{5+\t}{\t-0.5}}
    \foreach \t in {7,...,9}{\MErd{18-\t}{\t-0.5}}
  \end{tikzpicture}
  = 2^{-(2l-1+t)}\left[1+(1-p)^t\right]^{2l-1}
  \begin{tikzpicture}[baseline={([yshift=-0.6ex]current bounding box.center)},scale=0.5]
    \foreach \x in {6}{
      \foreach \y in {2,4,6,8}{
        \bendLud{\x+2}{\y+0.5}{\y+1.5}
        \obsWiggly{\x+1.8}{\y+1}
      }
    }
    \foreach \t in {4,6,8}{\propS{8.5}{\t}{colU}{black}}
    \foreach \t in {5,7}{\propS{9.5}{\t}{colU}{black}}
    \foreach \t in {6}{\propS{10.5}{\t}{colU}{black}}
    \foreach \t in {3,...,6}{\MEld{5+\t}{\t-0.5}}
    \foreach \t in {7,...,10}{\MErd{18-\t}{\t-0.5}}
  \end{tikzpicture}\\
    &=:
    \left(\frac{1+(1-p)^t}{2}\right)^{2l-1}
  T_{t}(p),
\end{aligned}
\end{equation}
where we introduced $T_t(p)$ to denote the value of the triangular diagram on
the right multiplied by $2^{-t}$.  Analogously to before, we are not able to
exactly evaluate $T_{t}(p)$ for generic $p$, but we can straightforwardly bound
it from above and below by its values in the two deterministic points,
$T_{t}(0)$ and $T_{t}(1)$. The lower bound is obtained by using the positivity of $p$ and $1-p$ to show
\begin{equation}
\label{eq:SMupperb}
  T_{t}(p)\ge
  2^{-t}
  \begin{tikzpicture}[baseline={([yshift=-0.6ex]current bounding box.center)},scale=0.5]
    \foreach \x in {6}{
      \foreach \y in {2,4,6,8}{
        \bendLud{\x+2}{\y+0.5}{\y+1.5}
        \obsZero{\x+1.8}{\y+1}
      }
    }
    \foreach \t in {8,7,6}{
      \tgridLine{8.5+8-\t}{\t}{8-\t+9.25}{\t+0.75}
      \MErd{8-\t+9.25}{\t+0.75}
    }
    \foreach \t in {4,...,6}{
      \tgridLine{4.5+\t}{\t}{5.25+\t}{\t-0.75}
      \MEld{5.25+\t}{\t-0.75}
    }
    \foreach \t in {4,6,8}{\propS{8.5}{\t}{colU}{black}}
    \foreach \t in {5,7}{\propS{9.5}{\t}{colU}{black}}
    \foreach \t in {6}{\propS{10.5}{\t}{colU}{black}}
    \foreach \t in {3}{\MEld{5+\t}{\t-0.5}}
    \foreach \t in {10}{\MErd{18-\t}{\t-0.5}}
    \foreach \t in {4,...,6}{\obsZero{5+\t}{\t-0.5}}
    \foreach \t in {7,...,9}{\obsZero{18-\t}{\t-0.5}}
    \foreach \t in {4,5}{\obsZero{5+\t}{\t+0.5}}
    \foreach \t in {4,5}{\obsZero{5+\t}{\t+1.5}}
    \foreach \t in {4}{\obsZero{5+\t}{\t+2.5}}
    \foreach \t in {4}{\obsZero{5+\t}{\t+3.5}}
  \end{tikzpicture}
  =2^{-t}=T_{t}(1).
\end{equation}
To find the upper bound we use that the diagonal transformation in~\eqref{eq:defInactiveP} together with the projector to the state $1$ sum up to the identity,
\begin{equation}
  \begin{tikzpicture}[baseline={([yshift=-0.6ex]current bounding box.center)},scale=0.5]
    \tgridLine{0}{-0.5}{0}{0.5}
    \obsWiggly{0}{0}
  \end{tikzpicture}+
  p\,
  \begin{tikzpicture}[baseline={([yshift=-0.6ex]current bounding box.center)},scale=0.5]
    \tgridLine{0}{-0.5}{0}{0.5}
    \obsOne{0}{0}
  \end{tikzpicture}=
  \begin{tikzpicture}[baseline={([yshift=-0.6ex]current bounding box.center)},scale=0.5]
    \tgridLine{0}{-0.5}{0}{0.5}
  \end{tikzpicture},
\end{equation}
which -- together with the positivity of all the entries of the local gate -- gives
\begin{equation}
\label{eq:SMlowerb}
  T_{t}(p)\le 
  2^{-t}
  \begin{tikzpicture}[baseline={([yshift=-0.6ex]current bounding box.center)},scale=0.5]
    \foreach \x in {6}{
      \foreach \y in {2,4,6,8}{
        \bendLud{\x+2}{\y+0.5}{\y+1.5}
      }
    }
    \foreach \t in {4,6,8}{\propS{8.5}{\t}{colU}{black}}
    \foreach \t in {5,7}{\propS{9.5}{\t}{colU}{black}}
    \foreach \t in {6}{\propS{10.5}{\t}{colU}{black}}
    \foreach \t in {3,...,6}{\MEld{5+\t}{\t-0.5}}
    \foreach \t in {7,...,10}{\MErd{18-\t}{\t-0.5}}
  \end{tikzpicture}=1=T_{t}(0).
\end{equation}

In summary, we have the following bounds on the $p$-dependent partition sum,
\begin{equation}\label{eq:BoundsZinfty}
  2^{-t}
  \left(\frac{1+(1-p)^t}{2}\right)^{2l-1}
  \le
  \lim_{s\to\infty} Z_{l,t}(p,s)
  \le
  \left(\frac{1+(1-p)^t}{2}\right)^{2l-1}.
\end{equation}
In particular, this means that the partition sum in the $s\to\infty$ limit is in general larger than at $p=1$,
\begin{equation}
  \lim_{s\to\infty} Z_{l,t}(p,s)\ge
  \lim_{s\to\infty} Z_{l,t}(1,s).
\end{equation}

The above bounds hold for any $p$, and are tight, since they are saturated
by the two deterministic limits of the model. However, we can find an alternative
lower bound that is tighter for $p\le \frac{1}{2}$. By decomposing the
transformation~\eqref{eq:defInactiveP} as 
\begin{equation}
\label{eq:bluedot}
  \begin{tikzpicture}[baseline={([yshift=-0.6ex]current bounding box.center)},scale=0.5]
    \tgridLine{0}{-0.5}{0}{0.5}
    \obsWiggly{0}{0}
  \end{tikzpicture}=
  (1-p)\,
  \begin{tikzpicture}[baseline={([yshift=-0.6ex]current bounding box.center)},scale=0.5]
    \tgridLine{0}{-0.5}{0}{0.5}
  \end{tikzpicture}+
  p\,
  \begin{tikzpicture}[baseline={([yshift=-0.6ex]current bounding box.center)},scale=0.5]
    \tgridLine{0}{-0.5}{0}{0.5}
    \obsZero{0}{0}
  \end{tikzpicture},
\end{equation}
we have
\begin{equation}
  T_t(p)\ge (1-p)^t T_t(0)=(1-p)^t.
\end{equation}
This gives 
\begin{equation}\label{eq:StricterLowerBoundZinfty}
  \lim_{s\to\infty} Z_{l,t}(p,s)\ge
  \left(\frac{1+(1-p)^t}{2}\right)^{2l-1}(1-p)^t.
\end{equation}

\subsection{Phase transition in the dynamical large deviations}
The occurrence of the phase transition is a consequence of different scaling of
$Z_{l,t}(p,s)$ with the subregion size in the two limits. To show this, we
define the cumulant generating function $F_{l,t}(p,s)$, and its asymptotic
value $F(p,s)$ as
\begin{equation}
  F_{l,t}(p,s)=\frac{1}{l t} \log Z_{l,t}(p,s),\qquad
  F(p,s)=\lim_{l\to\infty} F_{l,l\xi}(p,s),
\end{equation}
where we introduced $\xi$ to represent the asymptotic ratio between the two
sizes of the space-time region, $\xi=t/l$.  Plugging in the results above we
have a well-behaved expansion around $s\approx 0$,
\begin{equation}\label{eq:AsymptoticScaling1}
    F(p,s)=-p s + \frac{1}{2} p s^2 (1+A_p) + \mathcal{O}(s^3),\qquad
   -\min\{1,\frac{3 p}{2}\} \le A_p \le \frac{p}{2},
\end{equation}
where we introduced $A_p$ to denote a $p$-dependent constant that we can bound
as shown above.  On the other hand, in the limit $t\to \infty$ the rescaled
cumulant generating function vanishes,
\begin{equation}\label{eq:AsymptoticScaling2}
    \lim_{s\to\infty} F(p,s) = 0.
\end{equation}
We now note that for finite $l,t$ by definition $F_{l,t}(p,s)$ is analytic and
\emph{convex} \cite{touchette2009the-large}. Combining this with the above
asymptotic scaling we have that in the thermodynamic limit the first derivative
must become discontinuous, and the model exhibits a first-order phase
transition.

\subsection{Trotter limit}
In the appropriate scaling limit of large $t$ and small $p$, the stochastic Floquet-East
model reproduces the dynamics of the standard continuous-time stochastic
East model~\cite{jackle1991a-hierarchically,ritort2003glassy}. Interestingly,
the above argument carries over to this limit as well. 

We start by introducing the continuous-time partition sum
$\mathcal{Z}_{l,t}(\Gamma,s)$, where $t$ now denotes \emph{real time} (and not
the number of time-steps), $l$ is as before the size of the subsystem of
interest, and $\Gamma$ is the rate with which the spins in the continuous-time
model flip. The dynamics of the continuous-time model is approximated by the
stochastic Floquet-East when for a small $\Delta$ we scale the number of
time-steps as $t/\Delta$, while the probability parameter goes as $\Delta
\Gamma$. This gives the following continuous-time limit of the partition sum
\begin{equation}
  \mathcal{Z}_{l,t}(\Gamma,s)=\lim_{\Delta\to 0}Z_{l,\frac{t}{\Delta}}(\Gamma \Delta,s).
\end{equation}
Using the small-$s$ expansion of the discrete-time partition sum we get,
\begin{equation}
  \mathcal{Z}_{l,t}(\Gamma,s)=1-s\Gamma l t + s^2 \frac{l t \Gamma}{2}(1+l t \Gamma) +
  \mathcal{O}(s^3),
\end{equation}
where we used that the upper and lower bounds in Eq.~\eqref{eq:boundSecondOrderZ}
coincide in the Trotter limit, and give us an exact expression.  Similarly, we
can plug-in the $s\to\infty$ bounds to obtain
\begin{equation}
  \mathrm{e}^{-\Gamma t} \left(\frac{1+\mathrm{e}^{-\Gamma t}}{2}\right)^{2l-1}
  \le\lim_{s\to\infty} \mathcal{Z}_{l,t}(\Gamma,s) \le
  \left(\frac{1+\mathrm{e}^{-\Gamma t}}{2}\right)^{2l-1}.
\end{equation}
Note that here we need to use the lower bound given by Eq.~\eqref{eq:StricterLowerBoundZinfty}, 
since for small $p$ it is stricter than that of Eq.~\eqref{eq:BoundsZinfty}.

We now have all the ingredients to understand the scaling of the
thermodynamic-limit of the continuous-time cumulant generating function,
\begin{equation}
  \mathcal{F}(\Gamma,s)=\lim_{l,t\to\infty} \frac{1}{l t} \log \mathcal{Z}_{l,t}(\Gamma,s).
\end{equation}
As before, for small $s$ it has a well-behaved expansion around $0$, while it goes to zero for $s\to\infty$. 
\begin{equation}
  \left.\mathcal{F}(\Gamma,s)\right|_{s\approx 0}
  =-\Gamma s +\frac{s^2}{2}\Gamma +\mathcal{O}(s^3),\qquad
  \lim_{s\to\infty} \mathcal{F}(\Gamma,s)=0,
\end{equation}
therefore our results prove the existence of the phase transition also in the Trotter limit.

\section{Distribution of empty space-time regions}\label{sec:zeros}
\begin{figure}
\includegraphics[width=14cm]{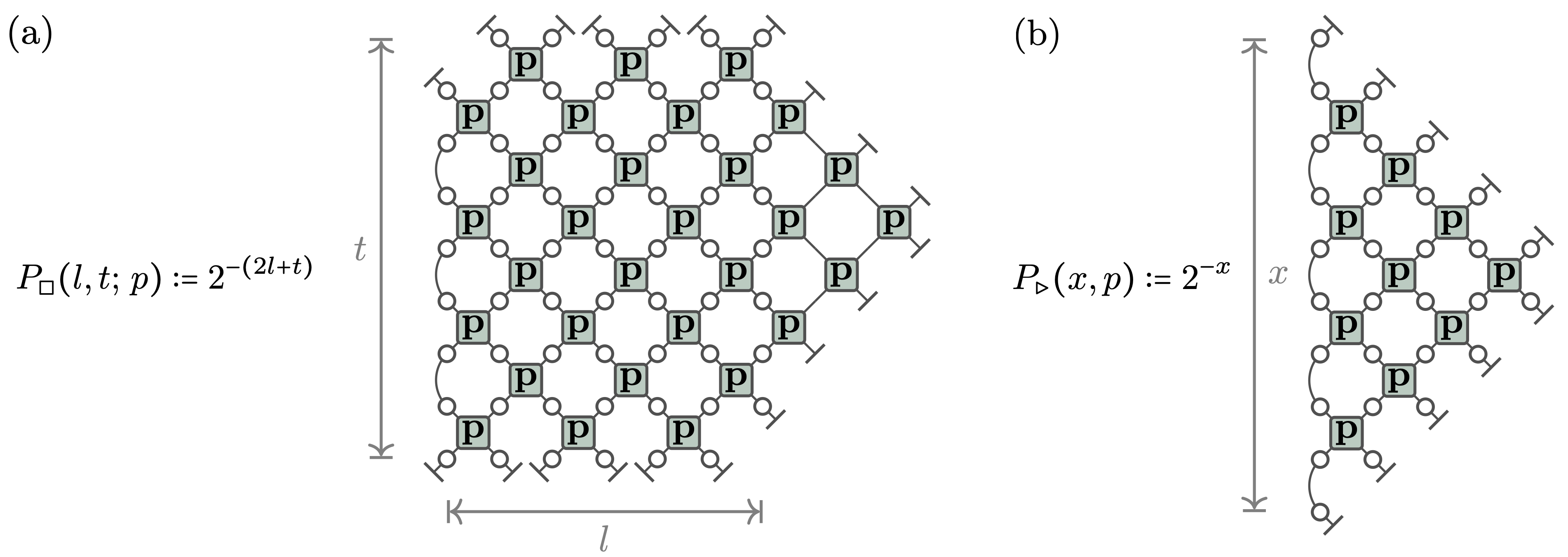}
  \caption{\label{fig:Fig3} (a) Probability of finding a region of states zero within a box of size $l \times t$. (b) Probability of finding a region of states zero within a equilateral triangle with base being of length $x$. }
\end{figure}
We now consider the probability $P_{\square}(l,t;p)$ of finding an empty region
of size $l \times t$ within the dynamics, see Fig.~\ref{fig:Fig3}(a) for a
precise definition.  In particular, we want to study how the probability
scales for large sizes. We assume $l$ and $t$ to be integer numbers for
simplicity, however, generalisations to semi-integers follow naturally, and do
not change the scaling. In~\cite{klobas2023exact}, the above probability has
been computed in the deterministic limit when $p=1$ and gives
\begin{equation}
\label{eq:Psquaredet}
      P_{\square}(l,t; \, 1)=   
  \frac{1}{2}   \lim_{s\to\infty} Z_{l,t}(1,s) =  2^{-(2l+t)} \,. 
\end{equation}
Thus, the probability $ P_{\square}(l,t; \, 1)$ manifestly scales with
perimeter. Furthermore, a rectangular condition of emptiness such as the one
displayed in Fig.~\ref{fig:Fig3}(a) implies a larger triangle surrounding the
box to be all in states zero. As a consequence we have the following equality,
\begin{equation}
\label{eq:squaretriang}
     P_{\square}(l,t; \, 1) = P_{\rhd}(2l+t,1) \,,
\end{equation}
where $P_{\rhd}$ is the probability of finding an empty region of triangular
shape, see Fig.~\ref{fig:Fig3}(b) for definition in the stochastic model. In
particular, the identity
\begin{equation} \label{eq:id0}
  \begin{tikzpicture}[baseline={([yshift=-0.6ex]current bounding box.center)},scale=0.5]
    \nctgridLine{0.75}{0.75}{-0.75}{-0.75}
    \nctgridLine{-0.75}{0.75}{0.75}{-0.75}
    \propS{0}{0}{colU}{black}
    \obsZero{-0.5}{0.5}
    \obsZero{0.5}{0.5}
    \obsZero{-0.5}{-0.5}
    \obsZero{0.5}{-0.5}
  \end{tikzpicture}
  =
  \begin{tikzpicture}[baseline={([yshift=-0.6ex]current bounding box.center)},scale=0.5]
    \bendRud{0}{-0.5}{0.5}
    \bendLud{1}{-0.5}{0.5}
    \obsZero{0.25}{0}
    \obsZero{0.75}{0}
  \end{tikzpicture}=  \begin{tikzpicture}[baseline={([yshift=-0.6ex]current bounding box.center)},scale=0.5]
    \nctgridLine{0.75}{0.75}{-0.75}{-0.75}
    \nctgridLine{-0.75}{0.75}{0.75}{-0.75}
    \prop{0}{0}{colU}
    \obsZero{-0.5}{0.5}
    \obsZero{0.5}{0.5}
    \obsZero{-0.5}{-0.5}
    \obsZero{0.5}{-0.5}
  \end{tikzpicture}\,,
\end{equation}
directly implies that the probability of an empty triangle does not depend on $p$,
\begin{equation}
\label{eq:emptytriangles}
  P_{\rhd}(x,p)=
  2^{-x}= P_{\rhd}(x,1).
\end{equation}

We need to stress that  Eq.~\eqref{eq:squaretriang} is a special property of
the DFE model and no longer holds for $p\neq1$, even
though the probability of a triangular hole does not depend on the precise
value of $p$. In particular, repeating the reasoning leading
to~\eqref{eq:emptytriangles}, one obtains the following expression for the
probability of a rectangular empty region for a generic value of $p$,
\begin{equation}  \label{eq:psquaredet0}
    P_{\square}(l,t; \, p)
  = 2^{-(2l+t)}
  \begin{tikzpicture}[baseline={([yshift=-0.6ex]current bounding box.center)},scale=0.5]
    \draw[thick,|<->|,gray] (7.25,2.5) -- (7.25,9.5) node[midway,left] {$t$};
    \foreach \x in {6}{
      \foreach \y in {2,4,6,8}{
        \bendLud{\x+2}{\y+0.5}{\y+1.5}
        \obsZero{\x+1.8}{\y+1}
      }
    }
    \foreach \t in {4,6,8}{\propS{8.5}{\t}{colU}{black}}
    \foreach \t in {5,7}{\propS{9.5}{\t}{colU}{black}}
    \foreach \t in {6}{\propS{10.5}{\t}{colU}{black}}
    \foreach \t in {3,...,6}{\MEld{5+\t}{\t-0.5}}
    \foreach \t in {7,...,10}{\MErd{18-\t}{\t-0.5}}
  \end{tikzpicture}=
  2^{-(2l+t)}
  \begin{tikzpicture}[baseline={([yshift=-0.6ex]current bounding box.center)},scale=0.5]
    \draw[thick,|<->|,gray] (7.25,2.5) -- (7.25,7.5) node[midway,left] {$t-1$};
    \foreach \x in {6}{
      \foreach \y in {2,4,6}{
        \bendLud{\x+2}{\y+0.5}{\y+1.5}
        \obsWiggly{\x+1.8}{\y+1}
      }
    }
     \foreach \t in {4,6}{\propS{8.5}{\t}{colU}{black}}
    \foreach \t in {5}{\propS{9.5}{\t}{colU}{black}}
    \foreach \t in {3,...,4,5}{\MEld{5+\t}{\t-0.5}}
    \foreach \t in {7,...,9}{\MErd{17-\t}{\t-0.5-1}}  
  \end{tikzpicture},
\end{equation}
with the r.h.s.\ straightforwardly following from the
definitions~\eqref{eq:defInactiveP}. The above partition sum is typically hard
to evaluate for general $t \in \mathbb{N}$, as the blue circles account for the
stochastic behaviour of the system, and the resulting graph is a complicated
polynomial function of $p$. However, it can be bounded by the two deterministic
limits through the very same argument presented above Eq.~\eqref{eq:SMlowerb}:
\begin{equation}
  P_{\square}(l,t \,; 1) = 2^{-2l-t}\leq P_{\square}(l,t \,; p) \leq 2^{-2l-1} 
  =  P_{\square}(l,t \,; 0) \,.
\end{equation}
It follows that the probability $P_{\square}(l,t \,; p)$ can \emph{at most}
scale with perimeter, and the triangular fluctuations characterising the DFE
model provide a lower bound for the stochastic probabilities.

\section{Distribution of generic space-time trajectories}\label{sec:traj}
One may wonder if the results of Sec.~\ref{sec:zeros} can be generalised to more general trajectories, and whether a probability of observing an empty region is somehow special compared to probabilities of observing more general space-time configuration. To address this question, we introduce the probability $P(\tau(l,t) ; p)$ of finding a particular trajectory $ \tau (l,t)$ inside a box of size $l \times t$. In the deterministic case, the probability $P(\tau(l,t) ; 1)$ equals the probability of having an empty region of the same size, provided that the trajectory $\tau(l,t)$ is allowed by the dynamics. We illustrate this by considering, for instance, a horizontal line of finite length $l$ of a random combination of states zero and one. The corresponding probability is 
\begin{equation} 
\label{eq:genericline}
P(\tau(l,1/2) ; 1)=  2^{-L}\,
  \begin{tikzpicture}[baseline={([yshift=-0.6ex]current bounding box.center)},scale=0.5]
    \foreach \x in {0,2,4}{
      \foreach \y in {0,2,4,6,8,10}
      {
        \prop{\x}{\y+1}{colU}
        \prop{\x+1}{\y}{colU}
      }
    }
    \foreach \x in {2,5}{\obsZero{\x-0.5}{4.5}}
    \foreach \x in {1,3,4}{\obsOne{\x-0.5}{4.5}}
    \foreach \x in {0,2,4}{
      \MEld{\x+1.5}{-0.5}
      \MErd{\x+0.5}{-0.5}
      \MErd{\x+0.5}{11.5}
      \MEld{\x-0.5}{11.5}
    }
    \draw[thick,gray,|<->|] (0.5,-0.875-0.5) --(4.5,-0.875-0.5) node [midway,below] {$l$};
  \end{tikzpicture}=  2^{-L}\,
  \begin{tikzpicture}[baseline={([yshift=-0.6ex]current bounding box.center)},scale=0.5]
    \foreach \x in {0,2,4}{
      \foreach \y in {0,2,4,6,8,10}
      {
        \prop{\x}{\y+1}{colU}
        \prop{\x+1}{\y}{colU}
      }
    }
    \foreach \x in {1,...,1}{\obsOne{\x-0.5}{9.5}}
    \foreach \x in {1,...,2}{\obsOne{\x-0.5}{8.5}}
    \foreach \x in {1,3}{\obsZero{\x-0.5}{7.5}}
    \foreach \x in {2}{\obsOne{\x-0.5}{7.5}}
    \foreach \x in {1,3}{\obsZero{\x-0.5}{6.5}}
    \foreach \x in {2,4}{\obsOne{\x-0.5}{6.5}}
    \foreach \x in {5}{\obsZero{\x-0.5}{5.5}}
    \foreach \x in {1,...,4}{\obsOne{\x-0.5}{5.5}}
    \foreach \x in {2,5}{\obsZero{\x-0.5}{4.5}}
    \foreach \x in {1,3,4}{\obsOne{\x-0.5}{4.5}}
    \foreach \x in {2,3}{\obsZero{\x-0.5}{3.5}}
    \foreach \x in {1,4}{\obsOne{\x-0.5}{3.5}}
    \foreach \x in {2,3}{\obsZero{\x-0.5}{2.5}}
    \foreach \x in {1}{\obsOne{\x-0.5}{2.5}}
    \foreach \x in {2}{\obsZero{\x-0.5}{1.5}}
    \foreach \x in {1}{\obsOne{\x-0.5}{1.5}}
    \foreach \x in {1,...,1}{\obsOne{\x-0.5}{0.5}}
    \foreach \x in {0,2,4}{
      \MEld{\x+1.5}{-0.5}
      \MErd{\x+0.5}{-0.5}
      \MErd{\x+0.5}{11.5}
      \MEld{\x-0.5}{11.5}
    }
  \draw[thick,gray,|<->|] (0.5,-0.875-0.5) --(4.5,-0.875-0.5) node [midway,below] {$l$};
  \end{tikzpicture}= 2^{-2l} \,
  \begin{tikzpicture}[baseline={([yshift=-0.6ex]current bounding box.center)},scale=0.5]
    \foreach \x in {0,...,4}{\nctgridLine{\x+0.5}{0.5+\x}{\x+0.75}{0.25+\x}}
    \foreach \x in {2,...,6}{\nctgridLine{\x+0.5-2}{11+0.5-\x}{\x+0.75-2}{11-\x+0.75}}
    \foreach \x in {0,...,4}{\MErd{\x+0.75}{11.75-2-\x}}
    \foreach \x in {0,...,4}{\MEld{\x+0.75}{\x+0.25}}
    \foreach \x in {0,2,...,8}{\bendLud{0.5}{\x+0.5}{\x+1.5}}
    \foreach \y in {2,4,...,8}{\prop{1}{\y}{colU}}
    \foreach \y in {3,5,...,7}{\prop{2}{\y}{colU}}
    \foreach \y in {4,6}{\prop{3}{\y}{colU}}
    \foreach \y in {5}{\prop{4}{\y}{colU}}
    \foreach \y in {0,1,2,3,4,5,8,9}{\obsOne{0.5}{\y+0.5}}
    \foreach \y in {6,7}{\obsZero{0.5}{\y+0.5}}
    \foreach \y in {1,...,4}{\obsZero{1.5}{\y+0.5}}
    \foreach \y in {5,...,8}{\obsOne{1.5}{\y+0.5}}
    \foreach \y in {2,3,6,7}{\obsZero{2.5}{\y+0.5}}
    \foreach \y in {4,5}{\obsOne{2.5}{\y+0.5}}
    \foreach \y in {3,...,6}{\obsOne{3.5}{\y+0.5}}
    \foreach \y in {4,...,5}{\obsZero{4.5}{\y+0.5}}
  \end{tikzpicture}=  P_{\rhd}(2l) \,.
\end{equation}
The above result reproduces the probability of finding an empty triangle of
base $2l$ in Eq.~\eqref{eq:emptytriangles}. Through
Eqs.~\eqref{eq:squaretriang} and \eqref{eq:genericline}, we can more generally
say that the probability $P(\tau(l,t) ; 1)$ of finding a box of size $l \times
t$ in a particular (deterministic) trajectory $ \tau(l,t)$ is
\begin{equation}
  P(\tau(l,t) ; 1 ) = 
  \begin{cases}
     P_{\square}(l,t; \, 1) &\text{for any allowed trajectory} \, \tau(l,t), \\
    0,& \text{otherwise}\,.
  \end{cases}
\end{equation}
Therefore the empty regions are not particularly special in the deterministic
case, and any allowed trajectory scales with perimeter.

The above statement no longer holds for the stochastic dynamics, where the
probability $P(\tau(l,t) ; p )$ strongly depends on the choice of trajectory
$\tau(l,t)$.  Indeed we can easily find trajectories $\tau(l,t)$ for which the
corresponding probabilities $P(\tau(l,t) ; p )$ scale with area rather than
perimeter, these include trajectories forbidden by the deterministic dynamics
such as the space-time region including all states one. The corresponding
probability is:
\begin{equation}
\label{eq:Pblacksquare}
    P_{\blacksquare}(l,t; \, p)=
  2^{-(2l+t)}
  \begin{tikzpicture}[baseline={([yshift=-0.6ex]current bounding box.center)},scale=0.5]
    \foreach \x in {2,4,6}{
      \nctgridLine{\x-0.25}{1.25}{\x+0.5}{2}
      \nctgridLine{\x+1.25}{1.25}{\x+0.5}{2}
      \MErd{\x-0.25}{1.25}
      \MEld{\x+1.25}{1.25}
    }
     \MEld{7+1.25}{1+1.25}
    \foreach \t in {3}{
      \tgridLine{4.5+\t}{\t}{5.25+\t}{\t-0.75}
    }
 \nctgridLine{2+0.5}{8}{2-0.25}{8.75}
\MEld{2-0.25}{8.75}
    \foreach \x in {2,4,6}{
      \nctgridLine{\x+1+0.5}{9}{\x+1+1.25}{9.75}
      \nctgridLine{\x+1+0.5}{9}{\x+1-0.25}{9.75}
      \MEld{\x+1-0.25}{9.75}
      \MErd{\x+1+1.25}{9.75}
    }
    \foreach \x in {2,4,6}{
      \foreach \y in {2,4,6,8}{
        \propS{\x+0.5}{\y}{colU}{black}
        \propS{\x+1.5}{\y+1}{colU}{black}
      }
    }
    \foreach \t in {2,4,6}{\bendLud{2}{\t+0.5}{\t+1.5}}
    \foreach \t in {4,6,8}{\propS{8.5}{\t}{colU}{black}}
    \foreach \t in {5,7}{\propS{9.5}{\t}{colU}{black}}
    \foreach \t in {6}{\propS{10.5}{\t}{colU}{black}}
    \foreach \t in {4,...,6}{\MEld{5+\t}{\t-0.5}}
    \foreach \t in {7,...,9}{\MErd{18-\t}{\t-0.5}}

      \foreach \x in {3,...,7}
    {\foreach \y in {1,...,9}{\obsOne{\x}{\y+0.5}}}
    \foreach \y in {1,...,8}{\obsOne{2}{\y+0.5}}

    \foreach \y in {2,...,9}{\obsOne{8}{\y+0.5}}
  \end{tikzpicture}
  = (1-p)^{2lt} \, 2^{-(2l+t)}
  \begin{tikzpicture}[baseline={([yshift=-0.6ex]current bounding box.center)},scale=0.5]
    \foreach \x in {6}{
      \foreach \y in {2,4,6,8}{
        \bendLud{\x+2}{\y+0.5}{\y+1.5}
        \obsOne{\x+1.8}{\y+1}
      }
    }
    \foreach \t in {4,6,8}{\propS{8.5}{\t}{colU}{black}}
    \foreach \t in {5,7}{\propS{9.5}{\t}{colU}{black}}
    \foreach \t in {6}{\propS{10.5}{\t}{colU}{black}}
    \foreach \t in {3,...,6}{\MEld{5+\t}{\t-0.5}}
    \foreach \t in {7,...,10}{\MErd{18-\t}{\t-0.5}}
  \end{tikzpicture}=
  (1-p)^{2lt}\, P_{\square}(l,t; \, p) \,,
\end{equation}
where $ P_{\square}(l,t; \, p) $ is given by~\eqref{eq:psquaredet0}. To arrive
to the second and third equality we respectively used the following two
observations,
\begin{equation}
    \begin{tikzpicture}[baseline={([yshift=-0.6ex]current bounding box.center)},scale=0.5]
      \nctgridLine{0.75}{0.75}{-0.75}{-0.75}
      \nctgridLine{-0.75}{0.75}{0.75}{-0.75}
      \propS{0}{0}{colU}{black}
      \obsOne{-0.5}{0.5}
      \obsOne{0.5}{0.5}
      \obsOne{-0.5}{-0.5}
     \obsOne{0.5}{-0.5}
    \end{tikzpicture}
    = (1-p)\,
  \begin{tikzpicture}[baseline={([yshift=-0.6ex]current bounding box.center)},scale=0.5]
    \bendRud{0}{-0.5}{0.5}
    \bendLud{1}{-0.5}{0.5}
    \obsOne{0.25}{0}
   \obsOne{0.75}{0}
  \end{tikzpicture}\,, \qquad  \begin{tikzpicture}[baseline={([yshift=-0.6ex]current bounding box.center)},scale=0.5]
    \foreach \x in {6}{
      \foreach \y in {2,4,6,8}{
        \bendLud{\x+2}{\y+0.5}{\y+1.5}
        \obsOne{\x+1.8}{\y+1}
      }
    }
    \foreach \t in {4,6,8}{\propS{8.5}{\t}{colU}{black}}
    \foreach \t in {5,7}{\propS{9.5}{\t}{colU}{black}}
    \foreach \t in {6}{\propS{10.5}{\t}{colU}{black}}
    \foreach \t in {3,...,6}{\MEld{5+\t}{\t-0.5}}
    \foreach \t in {7,...,10}{\MErd{18-\t}{\t-0.5}}
  \end{tikzpicture} =  \begin{tikzpicture}[baseline={([yshift=-0.6ex]current bounding box.center)},scale=0.5]
    \foreach \x in {6}{
      \foreach \y in {2,4,6,8}{
        \bendLud{\x+2}{\y+0.5}{\y+1.5}
        \obsZero{\x+1.8}{\y+1}
      }
    }
    \foreach \t in {4,6,8}{\propS{8.5}{\t}{colU}{black}}
    \foreach \t in {5,7}{\propS{9.5}{\t}{colU}{black}}
    \foreach \t in {6}{\propS{10.5}{\t}{colU}{black}}
    \foreach \t in {3,...,6}{\MEld{5+\t}{\t-0.5}}
    \foreach \t in {7,...,10}{\MErd{18-\t}{\t-0.5}}
  \end{tikzpicture} \,.
\end{equation} 
In particular, Eq.~\eqref{eq:Pblacksquare} implies that the probability $ P_{\blacksquare}(l,t; \, p)$ scales with area for any $p\neq 1$ and large $l$ and $t$. Note that the result above holds for any trajectory forbidden by the deterministic dynamics -- i.e., any trajectory for which we have $ P(\tau(l,t) ; 1 )=0$. \\

Let us now consider the probability of a generic trajectory $\tau(l,t)$, such
as the following,
\begin{equation}
  \label{eq:probgendef}
  P(\tau(l,t); \, p)=
  2^{-(2l+t)}
  \begin{tikzpicture}[baseline={([yshift=-0.6ex]current bounding box.center)},scale=0.5]
    \foreach \x in {2,4,6}{
      \nctgridLine{\x-0.25}{1.25}{\x+0.5}{2}
      \nctgridLine{\x+1.25}{1.25}{\x+0.5}{2}
      \MErd{\x-0.25}{1.25}
      \MEld{\x+1.25}{1.25}
    }
    \MEld{7+1.25}{1+1.25}
    \foreach \t in {3}{
      \tgridLine{4.5+\t}{\t}{5.25+\t}{\t-0.75}
    }
    \nctgridLine{2+0.5}{8}{2-0.25}{8.75}
    \MEld{2-0.25}{8.75}
    \foreach \x in {2,4,6}{
      \nctgridLine{\x+1+0.5}{9}{\x+1+1.25}{9.75}
      \nctgridLine{\x+1+0.5}{9}{\x+1-0.25}{9.75}
      \MEld{\x+1-0.25}{9.75}
      \MErd{\x+1+1.25}{9.75}
    }
    \foreach \x in {2,4,6}{
      \foreach \y in {2,4,6,8}{
        \propS{\x+0.5}{\y}{colU}{black}
        \propS{\x+1.5}{\y+1}{colU}{black}
      }
    }
    \foreach \t in {2,4,6}{\bendLud{2}{\t+0.5}{\t+1.5}}
    \foreach \t in {4,6,8}{\propS{8.5}{\t}{colU}{black}}
    \foreach \t in {5,7}{\propS{9.5}{\t}{colU}{black}}
    \foreach \t in {6}{\propS{10.5}{\t}{colU}{black}}
    \foreach \t in {4,...,6}{\MEld{5+\t}{\t-0.5}}
    \foreach \t in {7,...,9}{\MErd{18-\t}{\t-0.5}}

    \foreach \y in {1,...,8}{\obsOne{2}{\y+0.5}}

    \foreach \y in {1,...,9}{\obsZero{3}{\y+0.5}}

    \foreach \y in {2,...,9}{\obsZero{4}{\y+0.5}}
    \foreach \y in {1}{\obsOne{4}{\y+0.5}}

    \foreach \y in {5,...,9}{\obsZero{5}{\y+0.5}}
    \foreach \y in {1,...,4}{\obsOne{5}{\y+0.5}}

    \foreach \y in {1,2,3}{\obsZero{6}{\y+0.5}}
    \foreach \y in {4,...,9}{\obsOne{6}{\y+0.5}}

    \foreach \y in {1,2,7,8,9}{\obsZero{7}{\y+0.5}}
    \foreach \y in {3,...,6}{\obsOne{7}{\y+0.5}}    

    \foreach \y in {2,3,6,7}{\obsOne{8}{\y+0.5}}
    \foreach \y in {4,5,8,9}{\obsZero{8}{\y+0.5}}
  \end{tikzpicture} \,.
\end{equation}
The above probability always factorises into two contributions,  $P_1 (\tau(l,t); \, p)$ and $P_2 (\tau(l,t); \, p)$, scaling with area and perimeter, respectively. Our particular example can be rewritten as
\begin{equation}
  P(\tau(l,t); \, p)=
  P_1 (\tau(l,t); \, p)\,
  \;\;
  \underbrace{ 2^{-(2l+t)}
  \begin{tikzpicture}[baseline={([yshift=-0.6ex]current bounding box.center)},scale=0.5]
    \foreach \x in {6}{
      \foreach \y in {2,4,6,8}{
        \bendLud{\x+2}{\y+0.5}{\y+1.5}
      }
      \obsOne{\x+1.8}{2+1}
      \obsOne{\x+1.8}{6+1}
      \obsZero{\x+1.8}{4+1}
      \obsZero{\x+1.8}{8+1}
    }
    \foreach \t in {4,6,8}{\propS{8.5}{\t}{colU}{black}}
    \foreach \t in {5,7}{\propS{9.5}{\t}{colU}{black}}
    \foreach \t in {6}{\propS{10.5}{\t}{colU}{black}}
    \foreach \t in {3,...,6}{\MEld{5+\t}{\t-0.5}}
    \foreach \t in {7,...,10}{\MErd{18-\t}{\t-0.5}}
  \end{tikzpicture}}_{P_2 (\tau(l,t); \, p)}\,.
\end{equation}
where $P_1 (\tau(l,t); \, p)\,$ and the line of projectors onto zero and one in
the first column of $P_2 (\tau(l,t); \, p)\,$ are straightforwardly determined by the
trajectory within the box. In particular, we have that
\begin{equation}
    P_1 (\tau(l,t); \, p) = 
    p^{K} \, (1-p)^{2lt-K-J}\, ,
\end{equation}
where $K:=K(\tau(l,t))$ is the number of spin flips, and $J:=J(\tau(l,t))$ is the number of all the gates where the right spins are in the $0$ state (i.e., all the gates that would not flip in the $p=1$ limit). On the other hand, $P_2(\tau(l,t); \, p)$ can be bounded by using the following inequalities,
\begin{equation} 
2^{-t}\,
\begin{tikzpicture}[baseline={([yshift=-0.6ex]current bounding box.center)},scale=0.5]
\label{eq:in1}
    \foreach \x in {6}{
      \foreach \y in {2,4,6,8}{
        \bendLud{\x+2}{\y+0.5}{\y+1.5}
      }
      \obsOne{\x+1.8}{2+1}
      \obsOne{\x+1.8}{6+1}
      \obsZero{\x+1.8}{4+1}
      \obsZero{\x+1.8}{8+1}
    }
      \foreach \t in {8}{
      \tgridLine{8.5+8-\t}{\t}{8-\t+9.25}{\t+0.75}
      \MErd{8-\t+9.25}{\t+0.75}
    }
    \foreach \t in {4}{
      \tgridLine{4.5+\t}{\t}{5.25+\t}{\t-0.75}
      \MEld{5.25+\t}{\t-0.75}
    }
    
    \foreach \t in {4,6,8}{\propS{8.5}{\t}{colU}{black}}
    \foreach \t in {5,7}{\propS{9.5}{\t}{colU}{black}}
    \foreach \t in {6}{\propS{10.5}{\t}{colU}{black}}
    \foreach \t in {3,5,6}{\MEld{5+\t}{\t-0.5}} 
    \foreach \t in {7,8,10}{\MErd{18-\t}{\t-0.5}} 
    \foreach \y in {3,...,8}{\obsOne{9}{\y+0.5}}
    
  \end{tikzpicture} \leq \, 2^{-t}\,
  \begin{tikzpicture}[baseline={([yshift=-0.6ex]current bounding box.center)},scale=0.5]
    \foreach \x in {6}{
      \foreach \y in {2,4,6,8}{
        \bendLud{\x+2}{\y+0.5}{\y+1.5}
      }
      \obsOne{\x+1.8}{2+1}
      \obsOne{\x+1.8}{6+1}
      \obsZero{\x+1.8}{4+1}
      \obsZero{\x+1.8}{8+1}
    }
    \foreach \t in {4,6,8}{\propS{8.5}{\t}{colU}{black}}
    \foreach \t in {5,7}{\propS{9.5}{\t}{colU}{black}}
    \foreach \t in {6}{\propS{10.5}{\t}{colU}{black}}
    \foreach \t in {3,...,6}{\MEld{5+\t}{\t-0.5}}
    \foreach \t in {7,...,10}{\MErd{18-\t}{\t-0.5}}
  \end{tikzpicture} \leq \,2^{-t}\,  \begin{tikzpicture}[baseline={([yshift=-0.6ex]current bounding box.center)},scale=0.5]
    \foreach \x in {6}{
      \foreach \y in {2,4,6,8}{
        \bendLud{\x+2}{\y+0.5}{\y+1.5}
      }
    }
    \foreach \t in {4,6,8}{\propS{8.5}{\t}{colU}{black}}
    \foreach \t in {5,7}{\propS{9.5}{\t}{colU}{black}}
    \foreach \t in {6}{\propS{10.5}{\t}{colU}{black}}
    \foreach \t in {3,...,6}{\MEld{5+\t}{\t-0.5}}
    \foreach \t in {7,...,10}{\MErd{18-\t}{\t-0.5}}
  \end{tikzpicture} = 1\,.
\end{equation}
Above, the rightmost graph represents the sum over all the possible
trajectories within the triangle, while the leftmost graph is the sum over the
fewer trajectories of the central graph conditioned to having all ones in the
second column. Note that the inequalities above can be generalised for any
allowed trajectories and will always reproduce a non-trivial lower bound,
provided that $p\neq 0,1$.  Moreover, we can express the leftmost graph as
\begin{equation}
    2^{-t}\,
\begin{tikzpicture}[baseline={([yshift=-0.6ex]current bounding box.center)},scale=0.5]
    \foreach \x in {6}{
      \foreach \y in {2,4,6,8}{
        \bendLud{\x+2}{\y+0.5}{\y+1.5}
      }
      \obsOne{\x+1.8}{2+1}
      \obsOne{\x+1.8}{6+1}
      \obsZero{\x+1.8}{4+1}
      \obsZero{\x+1.8}{8+1}
    }
      \foreach \t in {8}{
      \tgridLine{8.5+8-\t}{\t}{8-\t+9.25}{\t+0.75}
      \MErd{8-\t+9.25}{\t+0.75}
    }
    \foreach \t in {4}{
      \tgridLine{4.5+\t}{\t}{5.25+\t}{\t-0.75}
      \MEld{5.25+\t}{\t-0.75}
    }
    
    \foreach \t in {4,6,8}{\propS{8.5}{\t}{colU}{black}}
    \foreach \t in {5,7}{\propS{9.5}{\t}{colU}{black}}
    \foreach \t in {6}{\propS{10.5}{\t}{colU}{black}}
    \foreach \t in {3,5,6}{\MEld{5+\t}{\t-0.5}} 
    \foreach \t in {7,8,10}{\MErd{18-\t}{\t-0.5}} 
    \foreach \y in {3,...,8}{\obsOne{9}{\y+0.5}}
  \end{tikzpicture}=
     2^{-t}\, p^{t-1/2-y} \,(1-p)^y \, 
  \begin{tikzpicture}[baseline={([yshift=-0.6ex]current bounding box.center)},scale=0.5]
    \foreach \x in {6}{
      \foreach \y in {2,4,6}{
        \bendLud{\x+2}{\y+0.5}{\y+1.5}
        \obsOne{\x+1.8}{\y+1}
      }
    }
     \foreach \t in {4,6}{\propS{8.5}{\t}{colU}{black}}
    \foreach \t in {5}{\propS{9.5}{\t}{colU}{black}}
    \foreach \t in {3,...,4,5}{\MEld{5+\t}{\t-0.5}}
    \foreach \t in {7,...,9}{\MErd{17-\t}{\t-0.5-1}}  
  \end{tikzpicture} = 2^{2l-1}\, p^{t-1/2-y} \,(1-p)^y \,  P_{\square}(l,t-1; \, p)\,,
\end{equation}
and the power $0\leq y \leq t$ depends on the particular trajectory. Thus, the inequalities in Eq.~\eqref{eq:in1} can be rewritten as 
\begin{equation}
    \frac{p^{t-1/2-y} \,(1-p)^y }{2} \, P_{\square}(l,t-1; \, p)\,\leq P_2 (\tau(l,t); \, p)  \leq 2^{-2l} \,.
\end{equation}
As a result, $P_2 (\tau(l,t); \, p)$ is at most exponential in $l$ and $t$, and thus scales with perimeter.  In contrast, the area-dependence comes from $P_1 (\tau(l,t); \, p)$, which can be written as:
\begin{equation}
    P_1 (\tau(l,t); \, p) = 2^{-\frac{2lt}{\log 2} \left(k \log p + (1-k-j) \log ( 1-p ) \right) }\,.
\end{equation}
where $k = K/ 2lt$ and $j = J/ 2lt$ are the densities of $k$ amd $j$ respectively. The dominant contribution can be related to the density of zero in the trajectory: we have that $  P (\tau(l,t); \, p)$ scales with perimeter i.e., $P_1 (\tau(l,t); \, p) \sim 1$ whenever empty regions dominate the trajectory i.e., $j \sim 1$ and $k \sim 0$, while it scales with area in any other case.

\section{Conclusions}\label{sec:conc}
In this paper, we have introduced and studied via exact tensor network methods a
stochastic generalisation of the classical deterministic Floquet-East model. We
proved that, just like the standard East model
\cite{garrahan2007dynamical,banuls2019using} or the deterministic Floquet-East
\cite{klobas2023exact}, with which it shares the local constraint of spin flips
only allowed in the vicinity of an excited nearest neighbours to one side (say,
to the left or East side), the stochastic Floquet-East model has a phase
transition at the large deviation level between an active and ergodic dynamical
phase and an inactive and non-ergodic dynamical phase. This dynamical
transition is first-order, meaning that even in the ergodic phase (favoured by
the vast majority of initial conditions) there are pronounced finite (in space
and time) fluctuations of inactivity that correspond to pre-transition effects:
for large enough inactive ``bubbles'' their log probability scales with
perimeter and not area, since it is more favourable to create a domain of the
inactive phase, paying only an interface cost. This is the same physics in
space-time of the celebrated hydrophobic effect in water
\cite{lum1999hydrophobicity,chandler2005interfaces} or the more general
orderphobic effect \cite{katira2016pre-transition}. We also showed that in an
appropriate limit the stochastic Floquet-East model corresponds to a
Trotterisation of the standard continuous-time East model, which means that our
results here provide an exact proof of dynamical hydrophobicity in that model,
cf`.\ Ref.~\cite{katira2018solvation}. These results highlight the usefulness
of exact tensor network methods for studying many-body stochastic systems. One
can think of many directions in which to expand the work here. These include
studying circuit versions of other kinetically constrained models,
generalisations to higher dimensions, and extentions to driven and
non-equilibrium circuit systems. 

\smallskip
\begin{acknowledgments}
\noindent
We acknowledge financial support from EPSRC Grant No.\ EP/V031201/1, and from The Leverhulme Trust through the Early Career Fellowship No.\ ECF-2022-324. K.\ K.\ warmly acknowledges the hospitality of the University of Ljubljana where this work was completed.
\end{acknowledgments}

\bibliographystyle{apsrev4-2}
\bibliography{bibliography}

\end{document}